# Optimal Needle Placement for Prostate Rotating-Shield Brachytherapy (RSBT)

Jirong Yi,[1] Quentin E. Adams,[2] Karolyn M. Hopfensperger,[3] Ryan T. Flynn,[2] Yusung Kim,[2] John M. Buatti,[2] Weiyu Xu,[1a] Xiaodong Wu[1,2a]

[1] *Department of Electrical and Computer Engineering, University of Iowa, Iowa City, IA 52242*

[2] *Department of Radiation Oncology, University of Iowa, 200 Hawkins Drive, Iowa City, IA 52242*

[3] *Huntsman Cancer Institute, University of Utah, Salt Lake City, UT 84112*

**ABSTRACT**

**Purpose:** To present an efficient NEEdle Position Optimization (NEEPO) algorithm for prostate rotating shield brachytherapy (RSBT). With RSBT, the increased flexibility beyond conventional high-dose-rate brachytherapy (HDR-BT) due to the partially shielded radiation source has been shown by Adams et al. in 2020 to enable improved urethra sparing (23.1% ±3.5%), enhanced dose escalation (29.9% ±3.0%), or both, with 20 needles without NEEPO-optimized positions. Within this regime of improved dosimetry, we propose in this work that the benefits of RSBT can be maintained while also reducing the number of needles needed for the delivery. The goal of NEEPO is to provide the capability to further increase the dosimetric benefit of RSBT and to minimize the number of needles needed to satisfy a dosimetric goal.

**Methods:** The NEEPO algorithm generates a needle pool for a given patient and then iteratively constructs a subset of needles from the pool based on relative needle importance as determined by total dwell times within needles. The NEEPO algorithm is based on a convex optimization formulation using a quadratic dosimetric penalty function, dwell time regularization by total variation, and a block sparsity regularization term to enable iterative removal of low-importance needles. RSBT treatment plans for 26 patients were generated using single fraction prescriptions with both dose escalation and urethra sparing goals, and compared to baseline HDR-BT treatment plans.

**Results:** For the dose escalation goal, NEEPO with 10 to 20 needles resulted in planning target volume (PTV) $D_{90\%}$ increases by 25.0% ±10.1% (applies to 10 needles with NEEPO) to 38.8% ±6.7% (applies to 20 needles with NEEPO), with a mean 30.2% ±3.8% increase (similar to 20 needles without NEEPO) achievable with 12 needles. Mean PTV $V_{200\%}$ and delivery time were reduced by 12.8% ±11.4% and 8.7% ±6.9%, respectively, when the number of needles ranged from 10 to 20. For the urethra sparing goal, NEEPO with 10 to 20 needles provided urethra $D_{10\%}$ reductions by 15.3% ± 8.4% to 27.8% ±4.5% with a mean of 18.7% ±3.1% (similar to 20 needles without NEEPO) occurring for 12 needles. Mean PTV $V_{200\%}$ and delivery time decreased by 52.0% ±10.0% and 17.8% ±7.8%, respectively, when the number of needles ranged from 10 to 20.

**Conclusions**: The NEEPO algorithm can improve RSBT dose escalation and urethra sparing, and substantially decrease the number of implanted needles needed to reach desired PTV $D_{90\%}$ and/or urethra $D_{10\%}$ levels. The PTV $V_{200\%}$ and treatment times increase as the needle number decreases, which would need to be considered in the needle reduction process, more so for urethra sparing than for dose escalation.

# I. INTRODUCTION

Prostate cancer is the most prevalent non-skin cancer among men in the United States.[1] Multi-fraction high-dose-rate brachytherapy (HDR-BT) is a localized prostate cancer treatment technique to achieve high biochemical control rates and low toxicity.[2] Due to the desire to reduce the number of needle implants and improve convenience for patients, single-fraction HDR-BT has been proposed, however, biochemical control rates for single-fraction HDR-BT have been substandard.[3,4,5] It is possible that biochemical control rates can be improved when the dose received by the planning target volume (PTV) is escalated given evidence of the effectiveness of dose escalation in treating prostate cancer.[6,7,8,9] The level of dose escalation achievable, however, is limited by the dose delivered to nearby organs at risk, especially the urethra.[10,11,12] For HDR-BT, the dose delivered to urethra can result in an increase in the risk of Grade 2 or higher GU toxicity and a decrease in the Expanded Prostate Index Composite urinary domain score.[10,13,14,15] This motivates the introduction of rotating shield brachytherapy (RSBT) for prostate cancer, which enables dose escalation to the planning target volume (PTV), urethra dose reduction, or both.[16,17,18]

To deliver RSBT for prostate cancer, a $^{169}$Yb source and delivery system was proposed to enable clinically practical delivery times with a commercially feasible isotope.[17,19] The mathematical models for RSBT treatment planning have usually been formulated as an optimization problem either in discrete form or continuous form, the optimization solvers also been developed correspondingly with combinatorial optimization or convex optimization.[20,21,22,23,24]

The placement or location of needles plays a critical role in determining the quality of prostate HDR-BT and RSBT treatment plans, with quality metrics including the maximally achievable PTV $D_{90\%}$ and the operational complexity such as the minimum number of needles required to achieve a certain PTV $D_{90\%}$ goal. Previously-proposed optimization techniques have used needle locations based on empirically-defined templates, resulting in needle locations that are feasible but not necessarily optimal for treatment planning.[17,25] In this work, an algorithm for optimizing the selection of needles for prostate RSBT is proposed, motivated by the desire for the highest possible PTV $D_{90\%}$ under the dose escalation goal or the lowest possible urethra $D_{10\%}$ under the urethra sparing goal. It is also desirable to minimize the number of needles for achieving a dosimetric goal. Our approach demonstrates the potential for improving upon previous approaches.[22,26,27,28] Concurrently, Wang et al. considered similar needle selection problem for traditional prostate HDR-BT (not RSBT, based on the $^{192}$Ir isotope), using a different algorithm to solve the problem.[29]

# II. MATERIALS AND METHODS

## II.A. Treatment planning

Twenty-six anonymous prostate cancer patients were considered, which were previously treated with HDR-BT. The same procedures for implementation of catheter and treatment planning as introduced by Adams et al. were followed after the optimal needles are selected.[17] The HDR-BT treatment plans were generated assuming a 10 Ci $^{192}$Ir Varian VariSource (Varian Medical Systems, Inc., Palo Alto, CA) radiation source, and the RSBT treatment plans were generated assuming a re-activatable 27 Ci $^{169}$Yb source.[19] The source activities were selected to ensure they both had the same dose rate in water at 1 cm off axis when unshielded. The same approach for dose delivery was assumed in the current work except that a needle selection technique was applied to select optimal needles for treatment planning.[17] Specifically, we use RSBT with NEEdle Position Optimization (NEEPO) to first select a set of optimal needles, and then use this set of optimal needles to generate optimal treatment plans. The key idea of generating treatment plans using the optimally selected needles is to solve a convex optimization problem where the objective function involves two terms, one for quantifying the difference between the generated treatment plan and the prescribed treatment plans, and the other for encouraging the smoothness of dose delivery.[17] The treatment plans obtained via RSBT NEEPO will then be compared with those obtained by using RSBT or HDR-BT only.

## II.B. Treatment planning with optimized needle selection

The NEEPO algorithm begins by generating a pool of needle positions from which optimized needle selection will be performed. To accomplish this, the maximum axial projection was first sampled at the original resolution as the CT image set. This image is then down-sampled into a 5 mm by 5 mm grid to reflect needle density constraints imposed on implantation of needles by commonly used prostate brachytherapy templates. Each down-sampled pixel position was defined as a potential programmatically generated needle position. Clinical needle positions were not considered as potential programmatic needle positions, as they instead form a clinical needle set. The set of dwell positions for clinical needles were sampled and the clinical needle with the most dwell positions was used as a basis needle to generate the dwell positions for each programmatic needle. The basis needle was assigned to each potential needle position by a linear shift, and the dwell positions from this basis needle were used for all the other needles. The PTV was decomposed into 4 quadrants of equal area on the maximum axial projection slice and the generated needles were each assigned to a quadrant.

Once the needle pool has been constructed, an optimization problem with block sparsity regularization is solved to determine the optimal set of needles to use. It is possible to preset the number of needles to be selected, $k$, and in this case, we can simply take the $k$ needles with the most dwell time after accounting for needle removal/redistribution as described in detail in this section. Once the optimal set of needles is selected, those needles are used to formulate a treatment planning optimization,[20] and solving this optimization problem will provide an optimal treatment plan with the optimally selected needles. We will evaluate the effects of needle selection primarily for RSBT with both the dose escalation goal and the urethra sparing goal. We will evaluate the needle selection techniques in two different scenarios: one where only the programmatic needles are for selection (clinical needles are excluded), and the other where both the clinical needles and the programmatic needles are used for selection.

In this section, we will formulate the needle selection as a convex optimization, and by solving it, we can decide which of the candidate needles should be used for treatment planning.

**(1) Optimal needle selection via block sparsity**

Assume there are a total $p \in Z$ needles in the needle pool for selection, and there are $n$ dwell positions for each needle. We now denote the dose rate matrix for the $u$-th needle by $D[u] \in R^{m \times n}$ where $m$ is the number of voxels of interest (VOIs), and the $i$-th row of $D[u]$ represents the dose rate for the $i$-th VOI due to the source from the $u$-th needle at different dwell positions. Then the dose rate matrix for all needles at all potential dwell positions can be written as $D = [D[1] \ D[2] \ \cdots \ D[p]] \in R^{m \times np}$. Define $t = [t[1]^T \ t[2]^T \ \cdots \ t[p]^T] \in R^{np}$ where each $t[u] \in R^n$ is the delivery time vector of the $u$-th needles at the corresponding $n$ dwell positions, and we can compute the dose for all VOIs $d \in R^m$ as $d = Dt$.

We propose to achieve optimal needle selection by solving

$$\min_{d,t} h(d) + \beta \|t\|_{TV} + \gamma \|t\|_{2,1} \quad \text{s.t.} \quad Dt = d, t \geq 0, \tag{1}$$

where the $\beta > 0$ and the $\gamma > 0$ are two positive regularization parameters. The function $h(d): R^m \to R$ is defined as

$$h(d) = \sum_{i=1}^{m} \left( \lambda_i^+ H(d_i - \hat{d}_i) + \lambda_i^- H(d_i - d_i) \right)(d_i - \hat{d}_i)^2 \tag{2}$$

Where $\lambda_i^+ \geq 0$ is the overdose penalty parameter, $\lambda_i^- \geq 0$ is the underdose penalty parameter, $d_i \geq 0$ is the prescribed dose for the $i$-th VOI, and $H(d_i - \hat{d}_i): R \to \{0,1\}$ is the Heaviside function defined as

$$H(d_i - \hat{d}_i) = \begin{cases} 1, & \text{if } d_i > \hat{d}_i \\ 0, & \text{if otherwise.} \end{cases} \tag{3}$$

The overdose penalty parameter $\lambda_i^+$ is positive if the $i$-th voxel belongs to OARs, or zero if the $i$-th voxel belongs to HR-CTV. The underdose parameter $\lambda_i^-$ is positive if the $i$-th voxel belongs to the HR-CTV, or zero if the $i$-th voxel belongs to the OARs. The function $\|t\|_{TV}: R^{np} \to R$ is defined as $\|t\|_{TV} = \sum_{i=1}^{np-1} |t_{i+1} - t_i|$, and by defining

$$L = \begin{bmatrix} -1 & 1 & 0 & 0 & \cdots & 0 & 0 \\ 0 & -1 & 1 & 0 & \cdots & 0 & 1 \\ \vdots & \vdots & \vdots & \vdots & \ddots & \vdots & \vdots \\ 0 & 0 & 0 & 0 & \cdots & -1 & 1 \end{bmatrix} \in R^{(np-1) \times np} \tag{4}$$

we obtain $\|t\|_{TV} = \|Lt\|_1$. The function $\|t\|_{2,1}: R^{np} \to R$ is defined as $\|t\|_{2,1} = \sum_{u=1}^{p} \|t[u]\|_2$, with $\|t[u]\|_2$ being the Euclidean norm, i.e., $\|t[u]\|_2 = \sqrt{\sum_{j=1}^{n} (t[u])_j^2}$. We used the $(t[u])_j$ to denote the $j$-th element of vector $t[u]$. The block sparsity regularization term $\|t\|_{2,1}$ enables iterative removal of low-importance needles.

**(2) Solve needle selection optimization via proximal operator graph solver**

In this section, we present an efficient algorithm for solving (1) via proximal operator graph solver (POGS).[31] Several algorithms based POGS have been previously presented solving treatment planning optimization problems.[17,20,21] The key is to convert our problem (1) to an optimization with graph constraints and derive the proximators of the objective functions. Once these are done, we follow the same ideas as we did in our previous work to iteratively find the optimal solution.[17,20,21]

We define $y_d$, $y_l$, and $y$ as $y_d = Dt \in R^m$, $y_t = Lt \in R^{np-1}$, $y = [y_d^T \ y_l^T]^T$. Then we can reformulate the problem (1) as

$$\min_{t,y} (g(y) + f(t)) \text{ s.t. } \begin{bmatrix} L \\ D \end{bmatrix} t = \begin{bmatrix} y_l \\ y_d \end{bmatrix} \tag{6}$$

where $g(y): R^{np-1+m} \to R$ is defined as $g(y) = (y_d - \tilde{d})^T \Lambda (y_d - \tilde{d}) + \beta \|y_l\|_1$ and $f(t): R^{np} \to R$ is defined as $f(t) = \gamma \|t\|_{2,1}, t \geq 0$.

From the definition of the proximator of a function $f(t)$ at a point $v$, i.e.,

$$\text{prox}_f(v) = \arg\min_{t \in \text{dom}(f)} \left( f(t) + \frac{\rho}{2} \|t - v\|^2 \right) \tag{7}$$

where $\rho > 0$ is a parameter and $\text{dom}(f)$ is the domain of $f$, we can obtain the proximator for the function $f(t)$ at $v \in R^{np}$ as

$$\text{prox}_f(v) = \arg\min_{t \geq 0} \left( \gamma \|t\|_{2,1} + \frac{\rho}{2} \|t - v\|^2 \right) = \begin{bmatrix} M \\ \max(\theta_u v[u], 0) \\ M \end{bmatrix}, \tag{8}$$

where

$$\theta_u = \begin{cases} \dfrac{\|v[u]\|_2 - \gamma/\rho}{\|v[u]\|_2}, & \|v[u]\|_2 > \gamma/\rho, \\ 0, & \text{otherwise.} \end{cases} \quad u = 1, 2, L \ p. \tag{9}$$

The proximator of $g(y)$ at $v = [v_l^T \ v_d^T]^T \in R^{np-1+m}$ with $v_l \in R^{np-1}$ and $v_d \in R^m$ can be obtained via

$$\text{prox}_g(v) = \arg\min_y \left( (y_d - \tilde{d})^T \Lambda (y_d - \tilde{d}) + \beta \|y_l\|_1 + \frac{\rho}{2} \|y - v\|_2^2 \right) = \begin{bmatrix} S_{\beta/\rho}(v_l) \\ y_d^* \end{bmatrix}, \tag{10}$$

where $S_{\beta/\rho}(v_l): {}^{np-1} \to {}^{np-1}$ is an element-wise function. The $k$-th element of $S_{\beta/\rho}(v_l)$ and the $i$-th element of $y_d^*$ are defined as

$$\left(S_{\beta/\rho}(v_l)\right)_k = \begin{cases} (v_l)_k - \beta/\rho, & \text{if } (v_l)_k > \beta/\rho, \\ 0, & \text{if } |(v_l)_k| < \beta/\rho, \\ (v_l)_k + \beta/\rho, & \text{if } (v_l)_k < -\beta/\rho, \end{cases} \quad , (y_d^*)_i = \begin{cases} \dfrac{2\lambda_i^+ (d\!\!\!/)_i + \rho(v_d)_i}{2\lambda_i^+ + \rho}, & (d\!\!\!/)_i < (v_d)_i, \\ (v_d)_i, & (d\!\!\!/)_i = (v_d)_i, \\ \dfrac{2\lambda_i^- (d\!\!\!/)_i + \rho(v_d)_i}{2\lambda_i^- + \rho}, & (d\!\!\!/)_i > (v_d)_i. \end{cases} \quad (11)$$

**(3) Optimal treatment planning with optimally selected needles**

Solving optimization problem (1) provides an initial estimate of $t \in R^{np}$, and the delivery time spent on the $u$-th needle is the sum of all the elements of $t[u] \in R^n$. The number of optimal needles allowed, $k(k \leq p)$, can be set by accepting $k$ needles which take up most of the total delivery time after taking into considerations needle redistribution.

More specifically, the optimized dwell times for the entire needle set were sorted by total dwell time for each needle and the number of desired needles for each case was selected. For the selected needles, the geometric quadrant indices were analyzed to ensure each quadrant contained a minimum of $1/8^{th}$ of the total number of selected needles. If a geometric quadrant contained an insufficient number of selected needles according to this criterion, then an iterative process was performed to ensure each quadrant was covered by sufficient needles. The needle with the lowest total dwell time in the selected group outside of the unfilled quadrant was replaced by the needle with the largest non-zero total dwell time from the post-optimization needle group that did not originally get selected. This process was repeated until the unfilled quadrant contained the required number of needles. The needle number criterion for each quadrant was re-checked and if any other needle quadrants did not meet criteria this process was repeated until all quadrants contained sufficient needle numbers.

Suppose after the selection of needles, the $k$ needles which achieved the highest total amount of delivery time are specified by index set $\{u_1, u_2, \ldots, u_k\} \subseteq \{1, 2, \ldots, p\}$. An optimized treatment plan with the selected needles is then obtained by solving (1) with $\gamma = 0$. This problem can be easily solved via the proximal operator graph solver.[17,20,21,30] After the dwell time solution is obtained, dose-volume metrics for clinical evaluation of treatment plans can be calculated.

## III. RESULTS

We use 26 patient cases with prostate tumor sizes ranging from 43.5 cm$^3$ to 92.7 cm$^3$, with average and standard deviation being 65.5 cm$^3$ and 14.6 cm$^3$, respectively. The number of clinical needles in RSBT for each patient is 20, while that in HDR-BT ranges from 20 to 24. For NEEPO, the number of programmatically generated needles for each patient ranged from 37 to 85 with average and standard deviation being 57.8 and 11.4.

In the first set of simulations, we present the optimized treatment plans without using needle selection techniques as baselines. We directly use all the clinical needles for treatment planning optimization, and the corresponding results are presented in Table 1. HDR-BT (Escalation) and RSBT Base. (Escalation) refer to the optimized treatment plans for HDR-BT without needle selection (HDR-BT) and RSBT without needle selection (RSBT), respectively, with the dose escalation goal. HDR-BT (Sparing) and RSBT Base. (Sparing) refer to the optimized treatment plans for HDR-BT and RSBT with the urethra sparing goal, respectively. From the results, we can see that under the dose escalation goal, RSBT can improve the PTV $D_{90\%}$ over HDR-BT by 30.2% ±3.8%, while reducing the number of needles by 10.5% ±6.3%, and the cost for these benefits is the longer treatment time, i.e., which was increased by 212.0% ±56.4%. Under urethra sparing goal, RSBT can reduce the urethra $D_{10\%}$ of HDR-BT by 23.2% ±2.3% and the number of needles by 10.5% ±6.3% at a cost of 139.5% ±43.6% increase in treatment time.

|  |  | HDR-BT (Escalation) | RSBT (Escalation) | Percentage | HDR-BT (Sparing) | RSBT (Sparing) | Percentage |
|---|---|---|---|---|---|---|---|
| PTV | $D_{90\%}$ (Gy) | 22.5 ±0.0 | 29.4 ±0.9 | 30.2% ±3.8% | 16.5 ±0.0 | 16.5 ±0.0 | 0.0% ±0.1% |
|  | $V_{100}$ (%) | 90.0 ±0.0 | 99.0 ±0.4 | 10.0% ±0.4% | 90.0 ±0.0 | 90.0 ±0.0 | 0.0% ±0.0% |
|  | $V_{150}$ (%) | 27.0 ±3.6 | 72.2 ±7.0 | 171.1% ±26.6% | 27.0 ±3.6 | 26.4 ±4.2 | -0.8% ±20.8% |
|  | $V_{200}$ (%) | 10.0 ±1.6 | 24.0 ±3.4 | 144.0% ±44.3% | 10.0 ±1.6 | 9.6 ±1.7 | -2.4% ±24.8% |
| Urethra | $D_{10\%}$ (Gy) | 21.2 ±0.3 | 21.2 ±0.3 | 0.0% ±0.1% | 15.5 ±0.2 | 11.9 ±0.4 | -23.2% ±2.3% |
|  | Mean (Gy) | 20.7 ±0.5 | 21.9 ±0.4 | 5.8% ±2.5% | 15.2 ±0.4 | 12.3 ±0.4 | -18.7% ±3.1% |
| Bladder | $D_{2cc}$ (Gy) | 14.9 ±1.3 | 16.6 ±1.7 | 11.5% ±5.6% | 10.9 ±0.9 | 9.4 ±1.1 | -14.3% ±4.7% |
| Rectum | $D_{2cc}$ (Gy) | 5.7 ±0.9 | 9.1 ±1.0 | 62.9% ±15.9% | 4.1 ±0.6 | 5.1 ±0.5 | -25.2% ±13.1% |
| Treatment Time (min) |  | 21.2 ±23.4 | 53.7 ±6.6 | 212.0% ±56.4% | 15.5 ±17.1 | 30.2 ±3.8 | 139.5% ±43.6% |
| # of Prescribed Needles |  | 22.5 ±1.5 | 20.0 ±0.0 | -10.5% ±6.3% | 22.5 ±1.5 | 20.0 ±0.0 | -10.5% ±6.3% |

Table 1: Optimal treatment plans without needle selection. The treatment plan statistics presented are averaged over all the patient cases. The percentage is the change for RSBT relative to HDR-BT.

In the second set of simulations, we apply the needle selection to obtain optimized treatment plans with optimally selected needles. The optimized treatment plans with the optimally selected needles in the case with only programmatic needles for selection are presented in Table 2, while those in the case with both clinical and programmatic needles for selection are presented in Table 3. By comparing the results in two different scenarios, we can see that they do not differ much, meaning that the incorporation of needle positions manually selected by physician experts is unnecessary for NEEPO. For example, under the dose escalation goal with 20 optimally selected needles, the p-value between PTV $D_{90\%}$ in Table 2 and that in Table 3 is 0.93, which means that the PTV $D_{90\%}$ in both Table 2 and 3 highly likely come from normal distributions with same mean and same variance.

|  | # of Selected Needles |  | 20 | 18 | 16 | 14 | 12 | 10 |
|---|---|---|---|---|---|---|---|---|
| NEEPO (Escalation) | PTV | $D_{90\%}$ (Gy) | 31.3±1.5 | 31.1±1.6 | 30.7±1.7 | 30.1±2.3 | 29.4±2.4 | 28.2±2.3 |
|  |  | $V_{100}$ (%) | 99.4±0.5 | 99.4±0.4 | 99.4±0.4 | 99.0±2.1 | 98.7±2.3 | 98.1±2.8 |
|  |  | $V_{150}$ (%) | 80.1±9.2 | 79.4±9.3 | 77.9±9.0 | 76.0±11.0 | 72.9±11.8 | 68.9±10.8 |
|  |  | $V_{200}$ (%) | 28.2±6.3 | 28.9±6.5 | 30.0±6.4 | 30.5±6.4 | 31.5±6.9 | 32.5±6.6 |
|  | Urethra | $D_{10\%}$ (Gy) | 21.2±0.8 | 21.3±0.7 | 21.5±0.8 | 21.7±0.7 | 22.0±0.6 | 22.3±0.5 |
|  |  | Mean (Gy) | 21.6±0.7 | 21.7±0.7 | 21.8±0.7 | 22.0±0.6 | 22.3±0.7 | 22.4±0.7 |
|  | Bladder | $D_{2cc}$ (Gy) | 14.8±1.7 | 14.9±1.7 | 15.1±1.8 | 15.4±1.5 | 15.6±1.3 | 15.7±1.4 |
|  | Rectum | $D_{2cc}$ (Gy) | 8.2±0.9 | 8.2±0.9 | 8.3±1.0 | 8.4±1.1 | 8.4±1.1 | 8.9±1.9 |
|  | Treatment Time (min) |  | 47.3±6.6 | 47.6±6.6 | 48.5±7.1 | 50.0±8.2 | 50.7±8.6 | 52.2±9.3 |
| NEEP | PTV | $D_{90\%}$ (Gy) | 16.5±0.0 | 16.5±0.0 | 16.5±0.0 | 16.5±0.0 | 16.5±0.0 | 16.5±0.0 |

| | | | | | | | |
|---|---|---|---|---|---|---|---|
| O (Sparing) | | $V_{100}$ (%) | 90.0±0.0 | 90.0±0.0 | 90.0±0.0 | 90.0±0.0 | 90.0±0.0 | 90.0±0.0 |
| | | $V_{150}$ (%) | 24.0±4.9 | 25.2±5.2 | 27.6±5.0 | 29.9±5.9 | 33.3±5.4 | 38.0±6.6 |
| | | $V_{200}$ (%) | 8.9±2.5 | 9.7±2.7 | 11.2±2.7 | 12.9±3.3 | 15.1±3.1 | 18.7±4.3 |
| | Urethra | $D_{10\%}$ (Gy) | 11.2±0.7 | 11.3±0.7 | 11.6±0.8 | 12.0±1.2 | 12.4±1.2 | 13.1±1.3 |
| | | Mean (Gy) | 11.4±0.6 | 11.5±0.6 | 11.8±0.7 | 12.1±1.0 | 12.6±1.1 | 13.2±1.1 |
| | Bladder | $D_{2cc}$ (Gy) | 7.9±1.2 | 8.0±1.2 | 8.2±1.2 | 8.5±1.4 | 8.8±1.3 | 9.3±1.3 |
| | Rectum | $D_{2cc}$ (Gy) | 4.3±0.6 | 4.4±0.6 | 4.5±0.6 | 4.6±0.7 | 4.8±0.8 | 5.3±1.5 |
| | Treatment Time (min) | | 25.1±4.4 | 25.4±4.5 | 26.3±4.9 | 27.7±6.7 | 28.8±6.9 | 31.0±7.5 |

Table 2: NEEPO with needle pool consisting of only the programmatic needles. The treatment plan statistics are averaged over all the patient cases.

| | # of Selected Needles | | 20 | 18 | 16 | 14 | 12 | 10 |
|---|---|---|---|---|---|---|---|---|
| NEEPO (Escalation) | PTV | $D_{90\%}$ (Gy) | 31.3±1.0 | 31.1±1.0 | 30.9±1.1 | 30.5±1.2 | 29.7±1.7 | 28.8±2.1 |
| | | $V_{100}$ (%) | 99.5±0.3 | 99.5±0.3 | 99.5±0.3 | 99.5±0.4 | 99.2±0.7 | 98.7±1.3 |
| | | $V_{150}$ (%) | 80.2±6.7 | 79.8±6.6 | 79.2±6.4 | 77.4±7.1 | 73.9±9.5 | 70.7±10.8 |
| | | $V_{200}$ (%) | 26.4±5.3 | 27.6±5.8 | 28.7±5.7 | 29.3±5.3 | 29.3±5.4 | 30.1±4.9 |
| | Urethra | $D_{10\%}$ (Gy) | 21.1±0.6 | 21.3±0.7 | 21.5±0.8 | 21.7±0.8 | 21.9±0.8 | 22.1±0.7 |
| | | Mean (Gy) | 21.5±0.6 | 21.7±0.7 | 21.9±0.7 | 22.1±0.8 | 22.2±0.8 | 22.4±0.8 |
| | Bladder | $D_{2cc}$ (Gy) | 14.7±1.7 | 14.9±1.7 | 15.2±1.7 | 15.4±1.7 | 15.7±1.7 | 15.8±1.6 |
| | Rectum | $D_{2cc}$ (Gy) | 8.1±0.9 | 8.1±0.9 | 8.3±0.9 | 8.4±1.0 | 8.5±1.0 | 8.6±1.2 |
| | Treatment Time (min) | | 46.8±6.3 | 47.3±6.3 | 48.2±6.5 | 49.1±7.4 | 50.4±8.0 | 51.4±8.3 |
| NEEPO (Sparing) | PTV | $D_{90\%}$ (Gy) | 16.5±0.0 | 16.5±0.0 | 16.5±0.0 | 16.5±0.0 | 16.5±0.0 | 16.5±0.0 |
| | | $V_{100}$ (%) | 90.0±0.0 | 90.0±0.0 | 90.0±0.0 | 90.0±0.0 | 90.0±0.0 | 90.0±0.0 |
| | | $V_{150}$ (%) | 22.5±3.6 | 24.2±3.9 | 26.0±4.0 | 27.8±4.6 | 30.1±5.2 | 33.7±6.4 |
| | | $V_{200}$ (%) | 8.0±1.8 | 9.1±1.9 | 10.3±2.2 | 11.8±2.7 | 13.5±3.0 | 15.9±4.1 |
| | Urethra | $D_{10\%}$ (Gy) | 11.1±0.5 | 11.3±0.5 | 11.5±0.6 | 11.8±0.6 | 12.2±0.8 | 12.7±1.0 |
| | | Mean (Gy) | 11.4±0.4 | 11.5±0.4 | 11.7±0.5 | 11.9±0.5 | 12.4±0.8 | 12.9±0.9 |
| | Bladder | $D_{2cc}$ (Gy) | 7.8±1.0 | 7.9±1.1 | 8.1±1.0 | 8.4±1.0 | 8.7±1.1 | 9.1±1.2 |
| | Rectum | $D_{2cc}$ (Gy) | 4.3±0.5 | 4.3±0.5 | 4.4±0.5 | 4.5±0.6 | 4.7±0.7 | 4.9±0.8 |
| | Treatment Time (min) | | 24.8±3.7 | 25.2±3.9 | 25.8±4.0 | 26.6±4.5 | 28.1±5.1 | 29.6±5.6 |
| Needles Composition | # of Clini. | | 5.8±2.6 | 5.3±2.3 | 4.7±2.1 | 4.3±2.2 | 3.8±1.9 | 3.3±1.8 |
| | # of Prog. | | 14.2±2.6 | 12.7±2.3 | 11.3±2.1 | 9.8±2.2 | 8.2±1.9 | 6.8±1.8 |

Table 3: NEEPO with needle pool being the combination of clinical needles and programmatic needles. The # of Pres. and the # of Prog. are the number of needles selected from the clinical needle set and the programmatic needle set, respectively. The treatment plan statistics are averaged over all the patient cases.

In the third set of simulations, we present in Figure 1 the boxplots of the PTV $D_{90\%}$ and the urethra $D_{10\%}$ in the treatment plans of all the patients obtained by NEEPO. The 'R' and 'H' refer to the RSBT and HDR-BT baselines without needle selection, respectively. We can see that there is a steady increase in the PTV $D_{90\%}$ when we increase the number of needles optimally selected, while the urethra $D_{10\%}$ decreases steadily.

In Figure 2, we present in the dose distribution maps and dose-volume histograms (DVHs) under the dose escalation goal for a particular patient (the patient # 19) to demonstrate the performance of our proposed method.

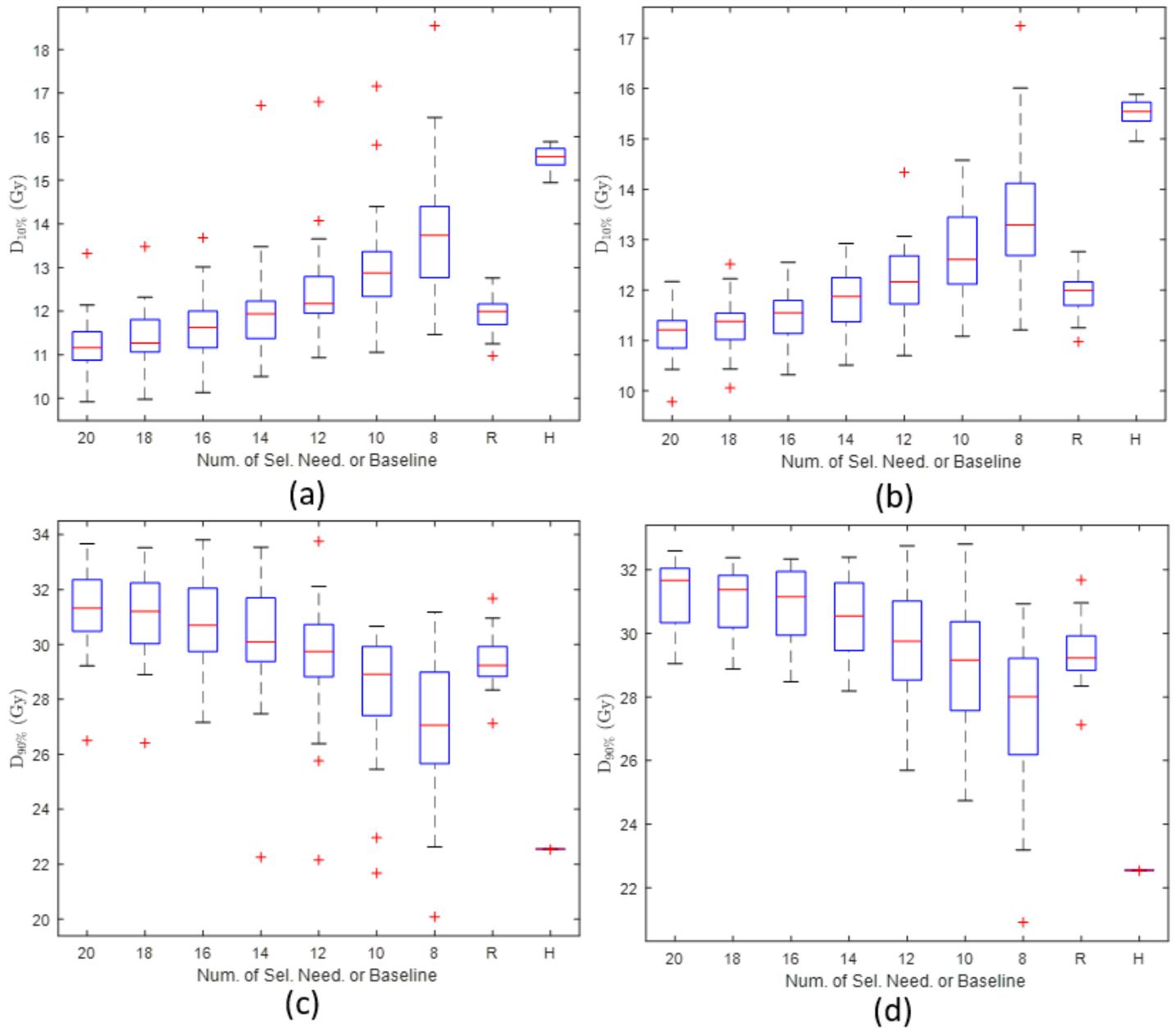

Figure 1: Boxplots of the urethra $D_{10\%}$ and the PTV $D_{90\%}$. (a) and (b) are obtained under the urethra sparing goal, while (c) and (d) are obtained under the dose escalation goal. (a) and (c) are obtained when only programmatic needles are used for selection while (b) and (d) are obtained when both clinical and programmatic needles are used for selection. The 'R' and 'H' refer to the RSBT and HDR-BT baselines without needle selection, respectively.

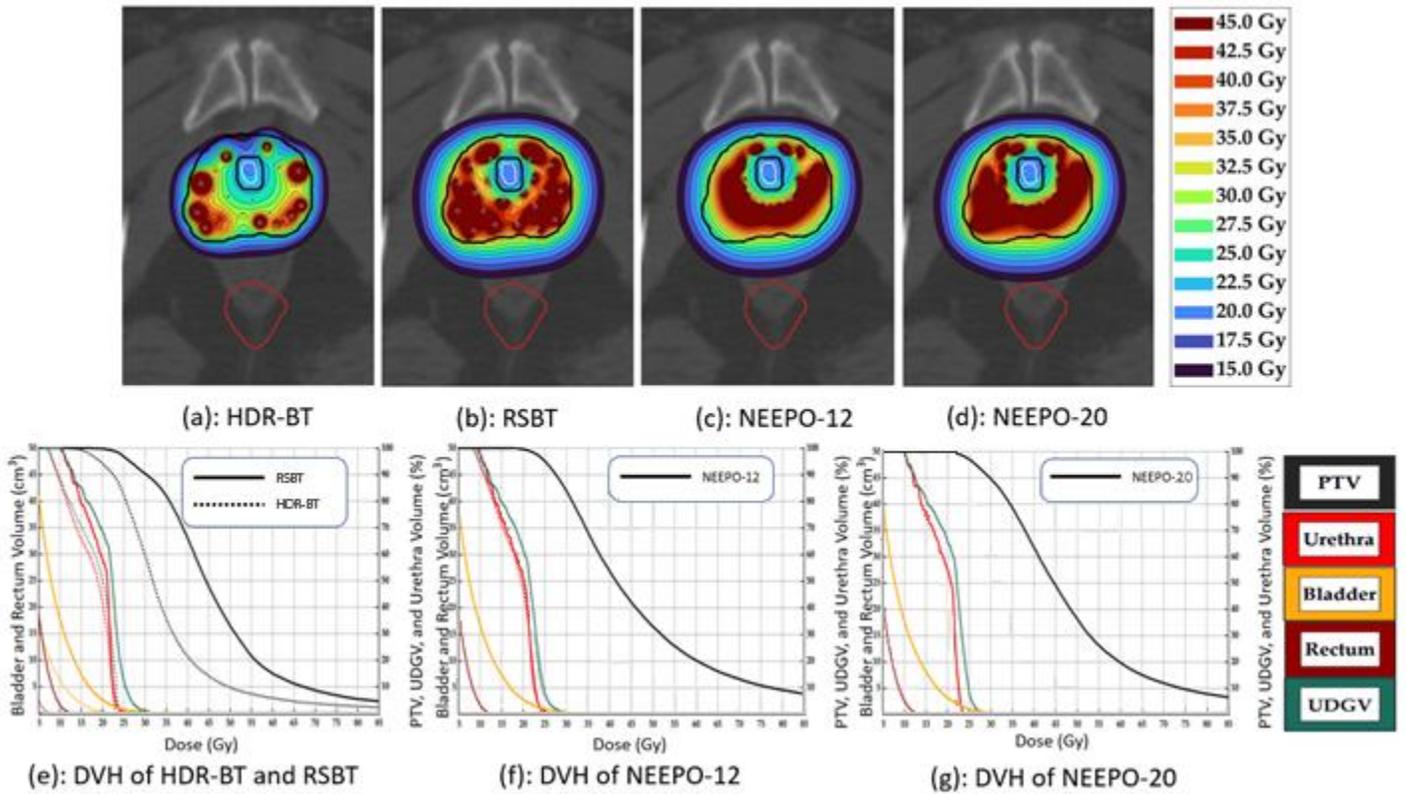

Figure 2: Dose distribution maps and DVHs of one patient case when only the programmatic needles are used for selection. In (a), HDR-BT achieves a PTV $D_{90\%}$ of 22.6 Gy with 22 needles in 14.2 minutes. In (b), RSBT achieves a PTV $D_{90\%}$ of 30.1 Gy using 20 needles in 47.1 minutes. In (c), NEEPO achieves a PTV $D_{90\%}$ of 29.9 Gy using 12 optimally selected needles in 42.9 minutes. In (d), NEEPO achieves PTV $D_{90\%}$ of 33.6 Gy using 20 optimally selected needles in 40.2 minutes. In (e), DVH of HDR-BT and RSBT. In (f), DVH of NEEPO with 12 optimally selected needles. In (g), DVH of NEEPO with 20 optimally selected needles.

## IV. DISCUSSION

The proposed needle selection is motivated by the following two aspects. Firstly, we aim to automatically select an optimal set of needles to achieve treatment plans with high quality, e.g., improving the PTV $D_{90\%}$ under the dose escalation goal or reducing the urethra $D_{10\%}$ for the purpose of urethral sparing. Our second aim is to reduce the operational complexity in clinical practice, e.g., the complexity can be reduced while fewer needles need to be placed.

From the simulation results in Table 2 and 3, we see that the optimized treatment plans with optimally selected needles remain almost the same in both the case where only programmatic needles are used for selection and the case where both the clinical and programmatic needles are used for selection. This implies that NEEPO is able to automatically select an optimal set of needles for prostate RSBT treatment planning with no or minimal intervention/guidance from physicians. Since adding the clinical needles to the needle pool for selection barely brings benefits, one can simply prescribe a single basis needle for dwell positions and apply it to the selected needles by NEEPO, which removes the heavy burden on physicians for manually prescribing best needles by trial-and-errors. Furthermore, as our future work, we can automatically generate the basis needle for dwell positions so that the system can be completely automated. We plan to develop an efficient algorithm to optimize the number of dwell positions associated with each needle and optimize the spacing between neighboring dwell positions.

A feasible way for optimal needle selection with concurrent optimization on dwell positions can be done by introducing an extra $L_1$ regularization term to the objective function, that is

$$\min_{d,t} h(d) + \beta \|t\|_{\text{TV}} + \gamma \|t\|_{2,1} + \delta \|t\|_1$$
$$\text{s.t.} \quad Dt = d,\ t \geq 0,$$

where the $\delta > 0$ is a parameter for controlling the tradeoff between the $L_1$ regularization and the other terms in the objective function. The $\|t\|_1$ is simply the sum of all the absolute value of all the elements of $t$.

From the results in Tables 1, 2, and 3, we can see that RSBT modality may always require a much longer delivery time than HDR-BT, i.e., roughly two times of that with HDR-BT is needed for NEEPO using 20 optimally selected needles. This is because a partially shielded $^{169}$Yb radiation source emits a lower photon energy per unit time into the PTV than an unshielded $^{192}$Ir source, even if both the $^{169}$Yb (27 Ci) and $^{192}$Ir (10 Ci) sources have the same unshielded dose rates at 1 cm lateral to the source in water, which was the case for the simulated sources in the current work. The total delivery time will slightly increase as the number of needles allowed for delivery decreases. However, as the proposed NEEPO method can effectively reduce the number of needles needed, it will be able to substantially shorten the needle placement time, which may certainty alleviate the negative effect of the longer treatment time.

# V. CONCLUSIONS

The NEEPO algorithm can improve RSBT dose escalation and urethra sparing, and substantially decrease the number of implanted needles needed to reach desired PTV $D_{90\%}$ and/or urethra $D_{10\%}$ levels. The PTV $V_{200\%}$ and delivery times increase as the needle number decreases would need to be considered in the needle reduction process, more so for urethra sparing than for dose escalation.

[a] **Author to whom correspondence should be addressed. Electronic email: weiyu-xu@uiowa.edu, xiaodong-wu@uiowa.edu.**